\documentclass[%
 aip,
 amsmath,amssymb,
reprint,%
author-numerical,%
]{revtex4-1}

\usepackage{dcolumn}% Align table columns on decimal point
\usepackage{bm}% bold math
\usepackage[mathlines]{lineno}% Enable numbering of text and display math
\usepackage{hjk}

\makeatletter
\def\@email#1#2{%
 \endgroup
 \patchcmd{\titleblock@produce}
  {\frontmatter@RRAPformat}
  {\frontmatter@RRAPformat{\produce@RRAP{*#1\href{mailto:#2}{#2}}}\frontmatter@RRAPformat}
  {}{}
}%
\makeatother

\newcommand{\rev}[1]{{\textcolor{black}{#1}}}
\newcommand{\xr}{{X}}
\newcommand{\yr}{{Y}}
\newcommand{\xp}{{x}}
\newcommand{\yp}{{y}}
\newcommand{\xc}{{x^{c}}}
\newcommand{\yc}{{y^{c}}}
\newcommand{\ex}{\tilde{x}}
\newcommand{\ey}{\tilde{y}}
\newcommand{\Ar}{A^{r}}
\newcommand{\Ap}{A^{p}}

\begin{document}

%\preprint{AIP/123-QED}

\title[]{Co-evolutionary control of \rev{a class of coupled mixed-feedback systems}}

\author{Luis Guillermo Venegas-Pineda}
\author{ Hildeberto Jardón-Kojakhmetov}
 \altaffiliation{CogniGron (Groningen Cognitive Systems and Materials Center), University of Groningen 
(Univ Groningen), Nijenborgh 4, NL-9747 AG Groningen, Netherlands.}
 \email{h.jardon.kojakhmetov@rug.nl}
 \affiliation{Bernoulli Institute for Mathematics, Computer Science and Artificial Intelligence, University of Groningen, Nijenborgh 9, 9747AG, Groningen, The Netherlands.}

\author{Ming Cao}
\affiliation{%
Engineering and Technology Institute Groningen, University of Groningen, Nijenborgh 4, 9747AG, Groningen, The Netherlands.%\\This line break forced% with \\
}%

\date{\today}% It is always \today, today,

\begin{abstract}
Oscillatory behavior is ubiquitous in many natural and engineered systems, often emerging through self-regulating mechanisms. In this paper, we address the challenge of stabilizing a desired oscillatory pattern in a networked system where neither the internal dynamics nor the interconnections can be changed. To achieve this, we propose two distinct control strategies. The first requires the full knowledge of the system generating the desired oscillatory pattern, while the second only needs local error information. In addition, the controllers are implemented as co-evolutionary, or adaptive, rules of some edges in an extended plant-controller network. We validate our approach in several insightful scenarios, including synchronization and systems with time-varying network structures.
\end{abstract}

\maketitle

\begin{quotation}
This work tackles the challenge of stabilizing a desired oscillatory pattern in a network with a fixed structure. We propose two distributed controllers to achieve this. These controllers are fundamentally different: the first is highly robust and eliminates nonlinearities but requires full knowledge of the reference system. The second, inspired by the neuromodulation of biological systems, only requires local error information. Moreover, the controllers are implemented as co-evolutionary rules for edges connecting an oscillatory node with the controlled system. This effectively results in an adaptive network where the adaptation rules shape the dynamics of the controlled nodes. We showcase the co-evolutionary controllers by exploring the synchronization of the controlled system, showing that our approach is independent of the network structure, and testing the performance of the controllers in time-varying networks. Our theoretical and numerical results show two distributed strategies to induce a desired oscillatory pattern by adaptively modulating node dynamics. 
\end{quotation}

\section{\label{sec:intro}Introduction}

Inducing and regulating a desired oscillatory behavior in networked systems is a fundamental problem in control theory, with ample relevance for both natural and engineered systems. Oscillations are ubiquitous in many real-world phenomena including industrial applications \cite{Ref:Wiesenfield1998, Ref:Pantaleone1998, Ref:Sajadi2022}, chemistry \cite{Ref:Taylor2009, Ref:Calugaru2020}, neural networks \cite{Ref:Uhlhaas2006, Ref:Shafiei2020}, psychology \cite{Ref:Correa2020, Ref:Symanski2022}, and biology \cite{Ref:Bick2020, Ref:Hauser2022}. More specifically, in biological systems, rhythmic signals govern important physiological processes, including the heartbeat's regulation, circadian rhythms, central pattern generators, and neuronal signaling \cite{tyson2008biological,Ref:Marder2001,Ref:Marder2012,Ref:Kiehn2016}. From the engineering side, several technologies also need such periodic signals, for example, to achieve the coordination of automation processes, including robotic locomotion \cite{ijspeert2008central,ijspeert2013dynamical}, or in the developments of synthetic biological systems \cite{potvin2016synchronous,Rosier,stricker2008fast}.

Several naturally occurring oscillatory patterns are self-regulated and emerge from the intrinsic dynamics of the system. In contrast, there are many scenarios, both natural and engineered, where external control is required to induce or modify oscillations in a desired way. This is, for instance, relevant in neuromorphic engineering\cite{Ref:Chicca2014}, where particular oscillatory patterns of biologically inspired neuronal networks are critical, for example, to mimic cognitive functions, coordinate sensorimotor functions, synchronize spiking neural networks, and enable efficient signal encoding, to name a few \cite{Angelidis2021ASC,Ref:Franklin2011,Paxon}. While our interests in this paper are mainly theoretical, designing controllers that can reliably induce desired oscillations could lead to advances in neuromorphic engineering and design, brain-inspired control systems, and adaptive network dynamics, among others.

In this paper, we design two controllers that impose a desired oscillatory pattern on a networked system \emph{without modifying its internal dynamics or interconnections}. Our focus is on a class of coupled mixed-feedback systems where the interplay of (fast) positive and (slow) negative feedback plays a key role in the generation of sustained oscillations (see \eqref{Eq:StaticNetwork} and further details below).

The manuscript is organized as follows: in Section \ref{sec:setup}, we motivate and describe the problem at hand and propose two kinds of controllers. The first controller, while robust, needs complete information from the reference. In contrast, the second one only requires local error information. Next, in Section \ref{sec:implementation}, we embed the designed controllers into a networked framework, implementing them as a co-evolutionary, or adaptive, rule of the edges connecting a node (the controller's node) to the nodes of the controlled system. We compare and evaluate the controllers' performance in three key scenarios: synchronization, arbitrary network structures, and time-varying network structures. We conclude in Section \ref{sec:conclusions}, where we also discuss potential extensions of our work.

{\bfseries{Notation:}} $\R$ and $\mathbb C$ denote the fields of real and complex numbers, respectively. We use $\sign$ for the sign function and $\cO$ to denote the big-Oh order symbol. An adjacency matrix is denoted by $A$ and a superscript $r$ (for ``reference'') or $p$ (for ``plant''), as in $\Ar$ and $\Ap$, are used whenever it is necessary to distinguish the adjacency matrices of the reference and of the plant. Further notation is clarified when necessary.

\section{\label{sec:setup}Problem setup and design of the controllers}

Our analysis is concerned with a class of dynamic networks given by the $2n$-dimensional system:
\begin{equation}\label{Eq:StaticNetwork}
    \begin{split}
        \ddt{\xp_i} &= -\xp_i-\yp_i+S(\alpha x_i + \beta \sum_{j=1}^nA_{ij}\xp_j)\\
        \ddt{\yp_i} &= \ve(\xp_i-\yp_i),
    \end{split}
\end{equation}
where $n\geq1$ is the number of nodes (oscillators) in the network, $0<\ve\ll1$ is a small parameter describing the timescale separation between the variables, and $S$ is a locally odd sigmoid function. This means, in particular, that the function $S:\R\to\R$ satisfies: a) $S(0)=0$, b) $\frac{\dd S}{\dd x}(x)>0$ $\forall x\in\R$, c) $\operatorname{argmax}(S'(x))=0$, and d) $S(x)=-S(-x)$ for $x\in U$ with $U$ a neighborhood of the origin. Moreover, $\alpha>0$, $\beta>0$ and $A=[A_{ij}]\in\R^{n\times n}$ is the adjacency matrix of the underlying graph. 

Model \eqref{Eq:StaticNetwork} is an example of a coupled mixed-feedback system \cite{Ref:Ribar2021,sepulchre2019control}. Each isolated node, namely
\begin{equation}\label{eq:mixed}
    \begin{split}
        \ddt{\xp_i} &= -\xp_i-\yp_i+S(\alpha x_i)\\
        \ddt{\yp_i} &= \ve(\xp_i-\yp_i),
    \end{split}
\end{equation}
exhibits a (fast) positive feedback through $S(\alpha x_i)$ (recall that $\alpha>0$) and (slow) negative feedback through the tendency of $\yp_i$ to follow $\xp_i$. 
\begin{remark}\label{rem:hopf}
    The nodes of \eqref{Eq:StaticNetwork}, which are given by \eqref{eq:mixed}, undergo a supercritical Hopf bifurcation of the origin for $\alpha=\alpha^*\coloneq\frac{1}{S'(0)}$. This means that the response of each isolated node converges to the origin for $\alpha<\alpha^*$, and to a limit cycle when $\alpha>\alpha^*$.
\end{remark}

As in \eqref{eq:mixed}, a key feature of mixed-feedback systems is the interplay between positive and negative feedback loops across different timescales, which is commonly observed in many biological models, particularly in neuronal dynamics. For example, it is argued \cite{sepulchre2019control} that mixed feedback is \emph{the} fundamental mechanism for excitability and explains how biological systems are able to exhibit sustained and robust oscillations \cite{che2023dominant,Ref:Tsai2008}. Moreover, mixed-feedback architectures have been recently exploited to develop control methods for neuromorphic systems \cite{Ref:Ribar2021}.

Our main motivation to consider systems given by \eqref{Eq:StaticNetwork} is the theory developed in Ref.~\onlinecite{Ref:Juarez2024}. Relevant to us is that they propose a procedure that allows one, for example, to design an adjacency matrix such that the output $\xp=(\xp_1,\ldots,\xp_n)\in\R^n$ corresponds to a \emph{rhythmic profile of the network} (defined as the \emph{inverse problem} in Ref.~\onlinecite{Ref:Juarez2024}). While brief, let us be more precise: a rhythmic profile is an $n$-tuple $(\sigma_1e^{\imath\phi_1},\ldots,\sigma_1e^{\imath\phi_1})\in\mathbb C^n$ where $\sigma_i\in\R$ represent amplitudes and $\phi_i\in[0,2\pi)$ phases. If for all $i=1,\ldots,n$, the solutions $\xp_i(t)$ of \eqref{Eq:StaticNetwork} converge to some periodic function with amplitude $\sigma_i$ and phase $\phi_i$ defined as \emph{oscillating function}\cite{Ref:Juarez2024}, then one says that the network is rhythmic. Assume further that $\sigma_1>0$ and that $\sigma_1\geq\sigma_i$, for all $i=2,\ldots,n$. The $n$-tuple $(1,\rho_2e^{\imath\theta_2},\ldots,\rho_ne^{\imath\theta_n})$, where $\rho_i=\frac{\sigma_i}{\sigma_1}$ and $\theta_i=\phi_i-\phi_1$ for all $i=2,\ldots,n$, is called \emph{a relative rhythmic profile}. By following the procedure summarized below, one can construct an adjacency matrix such that the solutions of \eqref{Eq:StaticNetwork} (locally) converge to a particular rhythmic profile:
\begin{enumerate}
    \item Let $\omega_x=(1,\rho_2e^{\imath\theta_2},\ldots,\rho_ne^{\imath\theta_n})$ with $\theta_i\neq\left\{ 0,\pi\right\}\mod2\pi$. 
    \item Pick $\mu_1=a+\imath b$ with $a,b>0$. Further choose $\mu_3,\ldots,\mu_n\in\R$ with $\mu_i<a$ for all $i=3,\ldots,n$. Define $D=\operatorname{diag}(\mu_1,\overline{\mu_1},\mu_3,\ldots,\mu_n)$.
    \item Find an \emph{invertible} matrix $Q=\begin{bmatrix}
        w_x & \overline{w_x} & B
    \end{bmatrix}\in\mathbb C^{n\times n}$, where $B\in\R^{n\times(n-2)}$.
    \item Define $A=QDQ^{-1}$.
\end{enumerate}

\begin{remark}\leavevmode
\begin{itemize}
    \item We refer to an adjacency matrix obtained from the above algorithm as \emph{rhythmic}.
    \item Once one picks $\mu_1$, one can then choose the parameters $\alpha$ and $\beta$ such that the origin of \eqref{Eq:StaticNetwork} is unstable, leading to the rhythmic profile through a Hopf bifurcation, recall Remark \ref{rem:hopf} and see Ref.~\onlinecite[Section VIII]{Ref:Juarez2024}.
    \item In Ref.~\onlinecite{Ref:Juarez2024}, the slow-fast as well as the mixed-feedback structure of \eqref{Eq:StaticNetwork} play a fundamental role. This structure will also be relevant for the controllers presented in sections \ref{sec:setup} and \ref{sec:implementation}. 
\end{itemize}
    
\end{remark}

We now consider the following problem: suppose we are given a reference rhythmic network and a plant network, both of type \eqref{Eq:StaticNetwork}. The adjacency matrix of the plant cannot be adjusted, but we want to render the plant rhythmic with respect to the reference's profile. Since the adjacency matrix of the plant is fixed, we shall design a controller that solves our problem.
\begin{remark}
    While our focus is theoretical, we notice that the described situation frequently appears in real-life scenarios: a) many neuromorphic devices have fixed hardwired interconnections but allow modulation through externally controllable nodes; b) deep brain stimulation consists of applying external electrical pulses to the (fixed) brain inducing a desired rhythmic activity. Similar challenges arise in robotics, where fixed mechanical structures coordinate with body motion, or the synchronization of power grids, where the underlying network remains unchanged while controllers regulate, for example, frequency stability.
\end{remark}

The control problem at hand is schematized in Figure \ref{Fig:Diagrams} and is described by the following system of equations:
\begin{equation}\label{Eq:ClosedLoop_General}
\begin{split}
    \ddt{\xr_i}&=-\xr_i-\yr_i+S\left(\alpha_r \xr_i + \beta_r\sum_{j=1}^n\Ar_{ij}\xr_j\right),\\
    \ddt{\yr_i}&=\ve(\xr_i-\yr_i),\\
    \ddt{\xp_i}&=-\xp_i-\yp_i+S\left(\alpha_p \xp_i + \beta_p\sum_{j=1}^n\Ap_{ij}\xp_j+u_i\right),\\
    \ddt{\yp_i}&=\ve(\xp_i-\yp_i),
\end{split}
\end{equation}
where $(\xr,\yr)=(\xr_1,\ldots,\xr_n,\yr_1,\ldots,\yr_n)\in\R^{2n}$ are the states of the reference, $(\xp,\yp)=(\xp_1,\ldots,\xp_n,\yp_1,\ldots,\yp_n)\in\R^{2n}$ are the states of the plant, $\Ar$, and $\Ap$ denote respectively the adjacency matrices of the reference and of the plant, and $u_i$ is the controller input into the $i$-th node of the plant. The parameters $\alpha_r>0$ and $\beta_r>0$ are chosen according to the procedure sketched in section \ref{sec:intro} and together with $\Ar$ render the reference network rhythmic. The parameters $\alpha_p>0$, $\beta_p>0$, and $\Ap$ of the plant do not need to lead to a rhythmic profile. We also mention that the controller $u_i$ is proposed to directly influence the plant's nodes in the same way as its internal connections and not as an independent mechanism due to the way the controller will be implemented in Section \ref{sec:implementation}. 

\begin{figure}[ht]
    \begin{tikzpicture}
        \node at (-2,0){
        \begin{tikzpicture}[scale=0.65]
        \node[circle,draw,fill=blue!30!white, minimum size=15pt,
              inner sep=0pt, outer sep=0pt] (a) at (10:2) {\footnotesize$3$};
        \node[circle,draw,fill=blue!30!white, minimum size=15pt,
              inner sep=0pt, outer sep=0pt] (b) at (80:2) {\footnotesize$2$};
        \node[circle,draw,fill=blue!30!white, minimum size=15pt,
              inner sep=0pt, outer sep=0pt] (c) at (170:2) {\footnotesize$1$};
        \node[circle,draw,fill=blue!30!white, minimum size=15pt,
              inner sep=0pt, outer sep=0pt] (d) at (240:2) {\footnotesize$n$};
        \node[circle,draw,fill=blue!30!white, minimum size=15pt,
              inner sep=0pt, outer sep=0pt] (e) at (300:2) {\footnotesize$i$};
        \draw[->](a)--(b);
        \draw[->](a)--(c);
        \draw[->](a)--(d);
        \draw[->](d)--(b);
        \draw[->](d)--(c);
        \draw[->](b)--(e);
        \draw[loosely dotted,very thick,shorten <= 0.25cm, shorten >= 0.25cm] (e)--(a);
        \draw[loosely dotted,very thick,shorten <= 0.15cm, shorten >= 0.15cm] (e)--(d);
    \end{tikzpicture}
        };
        \node at (2,0){
        \begin{tikzpicture}[scale=0.65]
        \node[circle,draw,fill=green!30!white, minimum size=15pt,
              inner sep=0pt, outer sep=0pt] (a) at (0,0) {\footnotesize$n$};
        \node[circle,draw,fill=green!30!white, minimum size=15pt,
              inner sep=0pt, outer sep=0pt] (b) at (-1.5,2) {\footnotesize$1$};
        \node[circle,draw,fill=green!30!white, minimum size=15pt,
              inner sep=0pt, outer sep=0pt] (c) at (0,4) {\footnotesize$2$};
        \node[circle,draw,fill=green!30!white, minimum size=15pt,
              inner sep=0pt, outer sep=0pt] (d) at (2,4) {\footnotesize$3$};
        \node[circle,draw,fill=green!30!white, minimum size=15pt,
              inner sep=0pt, outer sep=0pt] (e) at (2,0) {\footnotesize$i$};
        \node[circle,draw,fill=red!30!white, minimum size=15pt,
              inner sep=0pt, outer sep=0pt] (f) at (1,2) {\footnotesize$c$};
        \draw[->](a)--(b);
        \draw[->](c)--(a);
        \draw[->](c)--(b);
        \draw[->](c)--(d);
        \draw[->](b)--(e);
        \draw[->,red](f)--(a);
        \draw[->,red](f)--(b);
        \draw[->,red](f)--(c);
        \draw[->,red](f)--(d);
        \draw[->,red](f)--(e);
        \draw[loosely dotted,very thick,shorten <= 0.25cm, shorten >= 0.25cm] (e)--(a);
        \draw[loosely dotted,very thick,shorten <= 0.3cm, shorten >= 0.25cm] (e) to [bend right=45](d);
    \end{tikzpicture}
        };
        \node at (0,-3){
        \begin{tikzpicture}
            \node[rectangle,draw,fill=blue!30!white] (r) at (-1.75,0) {\footnotesize$R$};
            \node[circle, draw, minimum size=10pt, inner sep=0] (circ) {$ $};
            \node[rectangle,draw,fill=red!30!white] (c) at (1.75,0) {\footnotesize$C$};
            \node[rectangle,draw,fill=green!30!white] (p) at (3.5,0) {\footnotesize$P$};
            \draw[->,shorten <= 0.2cm] (p)--++(1,0);
            \draw[->](p)--++(0.5,0)--++(0,-1)--++(-4,0)--(circ);
            \draw[->](circ)--(c);
            \draw[->](c)--(p);
            \draw[->](r)--(circ);
            \node at (-.3,.2){\scriptsize$+$};
            \node at (.25,-.2){\scriptsize$-$};
            \node at (-0.925,0.25){\scriptsize$(
                \xr, \yr)$};
            \node at (0.75,0.25){\scriptsize$(
                \ex, \ey)$};
            \node at (2.65,0.25){\scriptsize$u$};
            \node at (4.75,0.25){\scriptsize$(
                \xp, \yp)$};
        \end{tikzpicture}
        };
    \end{tikzpicture}
    \caption{\label{Fig:Diagrams}Network (upper) and block diagram (lower) representations of our control problem \eqref{Eq:ClosedLoop_General}, in which elements of the reference, plant, and control are presented in blue, green, and red, respectively. For the networks, the effect of the controller on the plant is displayed in red, while the rest of the connections are provided for illustrative purposes only, as any topology is allowed in our methodology. For the block diagram, uppercase $(\xr,\yr)$ and lowercase $(\xp,\yp)$ represent the variables of the reference and of the plant, respectively, from which the error variables $\ex\coloneq\xr-\xp$, and $\ey\coloneq\yr-\yp$ are defined for our analysis.}

\end{figure}

Let us define the errors $(\ex,\ey)\in\R^{2n}$ by:
\begin{equation}
    \ex_i\coloneq\xr_i-\xp_i, \qquad \ey_i\coloneq\yr_i-\yp_i.
\end{equation}
The corresponding error dynamics read as
\begin{equation}\label{eq:error}
    \begin{split}
        \ddt{\ex_i} &= -\ex_i-\ey_i + S\left(\alpha_r \xr_i + \beta_r\sum_{j=1}^n\Ar_{ij}\xr_j\right) \\
        &\quad- S\left(\alpha_p \xp_i + \beta_p\sum_{j=1}^n\Ap_{ij}\xp_j+u_i\right)\\
        \ddt{\ey_i}&=\ve(\ex_i-\ey_i).
    \end{split}
\end{equation}
\begin{remark}
    The error system given in \eqref{eq:error} is not in closed form. We shall deal with this when we design the controllers in the following subsections.
\end{remark}

In the following, we propose two different kinds of controllers that render the origin of \eqref{eq:error} locally stable. The first one eliminates the sigmoidal nonlinearities, which results in a robust controller but requires complete knowledge of the reference. For comparison purposes, we propose an additional controller inspired by the synaptic modulation of neuronal systems. Such a controller does not require any knowledge of the reference. For clarity of our exposition, these two types of controllers are first provided without considering their implementation, see Propositions \ref{prop:u_ideal} and \ref{prop:u_hebbian}. After that, we propose a particular way of implementing them in a networked system where all nodes are of type \eqref{Eq:StaticNetwork}, including the controller. This will mean that the controller action is implemented via the adaptation of the weights connecting the controller node and the plant's nodes; see more details in Section \ref{sec:implementation}.

\subsection{General aspects of the simulations}\label{sec:generalities_simulations}
We describe some generalities about the simulations that we present below. When required, more details are provided, and all codes are available in Ref. \onlinecite{Jardon-Kojakhmetov_Code_for_Co-evolutionary_2025}. In our simulations, the function $S$ is chosen as the odd sigmoid $S(\cdot)=\tanh(\cdot)$, the small parameter $\ve$ is fixed at $\ve=\frac{1}{100}$, while many of the other parameters are chosen at random. When we say that a parameter is chosen ``at random'', we always mean it under a uniform distribution and specify the interval of possible values. The following steps align with the algorithm described in Section \ref{sec:intro}:
\begin{enumerate}
    \item The relative amplitudes $\rho_i$, $i=2,\ldots,n$, are chosen at random within the interval $(\frac{1}{3},1)$. The lower value $\frac{1}{3}$ is simply chosen so that the $i$-th relative amplitude is not too small in the simulations. The phases $\theta_i$, $i=2,\ldots,n$, are chosen at random within the interval $(0,2\pi)$. The probability of $\theta_i=\pi$ is zero.
    \item For the leading eigenvalue $\mu_1=a+\imath b$ we let $a=1$ and $b$ is randomly chosen within the interval $(\frac{1}{100},\frac{1}{10})$. The choice of $b$ is so that the period of the rhythmic profile, which is\cite{Ref:Juarez2024} $T\approx2\pi(\beta\Im(\mu_1))^{-1}$, is not too small. The rest of the eigenvalues $\mu_i\in\R$, $i=3,\ldots,n$, are chosen at random within the interval $(0,\frac{9}{10}a)$. This provides $D=\operatorname{diag}(\mu_1,\overline{\mu_1},\mu_3,\ldots,\mu_n)$.
    \item The complement matrix $B$ is a randomly generated sparse matrix with nonzero coefficients within the interval $(0,1)$. This matrix is generated until $Q=\begin{bmatrix}
        w_x & \overline{w_x} & B
    \end{bmatrix}$ is invertible.
    \item The adjacency matrices $\Ar$ and $\Ap$ are thus given via the previous algorithm as $A^\bullet=QDQ^{-1}$. Since the algorithm to obtain the adjacency matrices is mostly random, both matrices are, with probability $1$, different. Likewise, since $B$ is sparse and randomly generated, the topologies of the adjacency matrices are, with probability $1$, different. We emphasize that, from the results we present below, the plant's adjacency matrix $\Ap$ does not have to be rhythmic. To keep all systems (reference, plant, and controller) within the same context, we have opted to keep it rhythmic for the simulations.
\end{enumerate}

In addition, and due to the choice of $\mu_1$, $\beta_\bullet$ is randomly chosen within the interval $(0,1)$, and the corresponding $\alpha_\bullet$ is set as $\alpha_\bullet=1+\ve-\beta_\bullet+\frac{1}{100}$ (see Ref. \onlinecite[Theorem 1]{Ref:Juarez2024}). 

To better see the effect of the controllers, we present our simulations in the following way: for the first $300$ time units, the controller is off, and so we see the open-loop response. At $t=300$, the controller is turned on, and so from thereon, we see the closed-loop response. In our algorithms, this is realized by multiplying the controller by $l(t-t_c)$, where $l(x)=\frac{1}{1+e^{-x}}$ and $t_c=300$ is the time at which the controller is turned on. Since we chose some parameters at random, the plots we show are representative of several simulation runs.

\subsection{Elimination of nonlinearities}\label{sec:ideal}

The first controller we propose follows from the next very simple observation: suppose that the controller can eliminate the sigmoidal nonlinearities of \eqref{eq:error}. The resulting error system is linear
\begin{equation}\label{eq:error0}
    \begin{split}
        \ddt{\ex_i} &= -\ex_i-\ey_i,\\
        \ddt{\ey_i}&=\ve(\ex_i-\ey_i),
    \end{split}
\end{equation}
and the origin is globally exponentially stable. Hence, the ideal controller that achieves this can be simply computed as follows:
\begin{equation}\label{eq:controlfull}
    \begin{split}
        u_i &= -\alpha_p\xp_i-\beta_p\sum_{j=1}^n\Ap_{ij}\xp_j+\alpha_r\underbrace{(\ex_i+\xp_i)}_{\xr_i}\\
        &\quad+\beta_r\sum_{j=1}^n\Ar_{ij}\underbrace{(\ex_j+\xp_j)}_{\xr_j}\\
        &=\alpha_r\ex_i+(\alpha_r-\alpha_p)\xp_i\\
        &\quad+\sum_{j=1}^n\beta_r\Ar_{ij}\ex_j(\beta_r\Ar_{ij}-\beta_p\Ap_{ij})\xp_j.
    \end{split}
\end{equation}
For future reference, let
\begin{equation}\label{eq:Fi}
\begin{split}
    F_i&\coloneq\alpha_r\ex_i+(\alpha_r-\alpha_p)\xp_i\\
        &\quad+\sum_{j=1}^n\beta_r\Ar_{ij}\ex_j(\beta_r\Ar_{ij}-\beta_p\Ap_{ij})\xp_j.
\end{split}
\end{equation}

For our implementations in Section \ref{sec:implementation}, it will be convenient that the controller is of integral type. Hence, we propose the closed-loop system:
\begin{equation}\label{Eq:ClosedLoop_Ideal}
\begin{split}
    \ddt{\xr_i}&=-\xr_i-\yr_i+S\left(\alpha_r \xr_i + \beta_r\sum_{j=1}^n\Ar_{ij}\xr_j\right),\\
    \ddt{\yr_i}&=\ve(\xr_i-\yr_i),\\
    \ddt{\xp_i}&=-\xp_i-\yp_i+S\left(\alpha_p \xp_i + \beta_p\sum_{j=1}^n\Ap_{ij}\xp_j+u_i\right),\\
    \ddt{\yp_i}&=\ve(\xp_i-\yp_i),\\
    \ddt{u_i}&=k_i(F_i-u_i).
\end{split}
\end{equation}
Indeed, we have:
\begin{proposition}\label{prop:u_ideal}
    Consider \eqref{Eq:ClosedLoop_Ideal} with $F_i$ given by \eqref{eq:Fi}. If $k_i\geq k>0$ for all $i=1,\ldots,n$ and with $k$ sufficiently large, then $\lim_{t\to\infty}|\ex(t)|=\cO(\frac{1}{k})$.
\end{proposition}
\begin{proof}
    It suffices to let $k_1=\cdots=k_n=k$. For $k\gg1$ sufficiently large, the dynamics of $u_i$ evolve in a fast timescale. In the limit when $k\to\infty$, $F_i$ is constant, and the equilibrium of $\ddtau{u_i}=(F_i-u_i)$, where $\tau=kt$ is the fast time, is hyperbolic. This is equivalent to saying that the critical manifold associated with \eqref{Eq:ClosedLoop_Ideal}, namely $\cC\coloneq\left\{ u_i=F_i\right\}$ is normally hyperbolic \cite{kuehn2015multiple} and globally exponentially stable. The restriction of \eqref{Eq:ClosedLoop_Ideal} to the critical manifold leads precisely to the linear error dynamics \eqref{eq:error0}, whose origin is globally exponentially stable. Since the equilibria of both the fast and the slow dynamics are exponentially stable, the result follows from, for example, Fenichel's (or Tikhonov's) theorem \cite{kuehn2015multiple,kokotovic1999singular}.
\end{proof}

\begin{figure*}[htbp]
    \centering
    \begin{tikzpicture}
    \node at (0,0){
    \begin{tikzpicture}
    \node at (0,0){
    \includegraphics[]{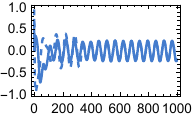}
    };
    \node at (0.5,-1.25) {\small$t$};
    \node[] at (0,1.2) {\small$\xp_1,\xr_1$};
    \end{tikzpicture}
    };
    \node at (3.5,0){
    \begin{tikzpicture}
    \node at (0,0){
    \includegraphics[]{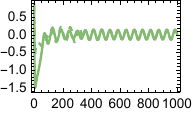}
    };
    \node at (0.5,-1.25) {\small$t$};
    \node[] at (0,1.2) {\small$\xp_2,\xr_2$};
    \end{tikzpicture}
    };
    \node at (7,0){
    \begin{tikzpicture}
    \node at (0,0){
    \includegraphics[]{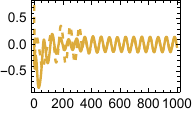}
    };
    \node at (0.5,-1.25) {\small$t$};
    \node[] at (0,1.2) {\small$\xp_3,\xr_3$};
    \end{tikzpicture}
    };
    \node at (10.5,0){
    \begin{tikzpicture}
    \node at (0,0){
    \includegraphics[]{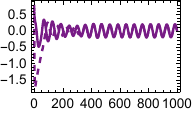}
    };
    \node at (0.5,-1.25) {\small$t$};
    \node[] at (0,1.2) {\small$\xp_4,\xr_4$};
    \end{tikzpicture}
    };
    \node at (0,-3){
    \begin{tikzpicture}
    \node at (0,0){
    \includegraphics[]{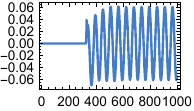}
    };
    \node at (0.5,-1.25) {\small$t$};
    \node[] at (0,1.2) {\small$u_1$};
    \end{tikzpicture}
    };
    \node at (3.5,-3){
    \begin{tikzpicture}
    \node at (0,0){
    \includegraphics[]{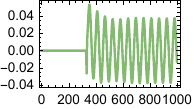}
    };
    \node at (0.5,-1.25) {\small$t$};
    \node[] at (0,1.2) {\small$u_2$};
    \end{tikzpicture}
    };
    \node at (7,-3){
    \begin{tikzpicture}
    \node at (0,0){
    \includegraphics[]{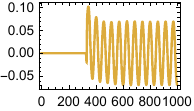}
    };
    \node at (0.5,-1.25) {\small$t$};
    \node[] at (0,1.2) {\small$u_3$};
    \end{tikzpicture}
    };
    \node at (10.5,-3){
    \begin{tikzpicture}
    \node at (0,0){
    \includegraphics[]{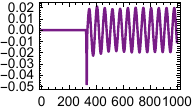}
    };
    \node at (0.5,-1.25) {\small$t$};
    \node[] at (0,1.2) {\small$u_4$};
    \end{tikzpicture}
    };
    %%%
    \node at (1.5,-7.25){
    \begin{tikzpicture}
    \node at (0,0){
    \includegraphics[]{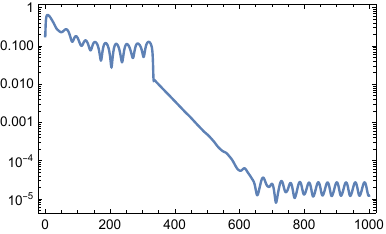}
    };
    \node at (0.25,2.25){\small$\frac{1}{n}\log|\ex|$};
    \node at (0.25,-2.25){\small$t$};
    \end{tikzpicture}
    };
    \node at (9,-7.25){
    \begin{tikzpicture}
    \node at (0,0){
    \includegraphics[]{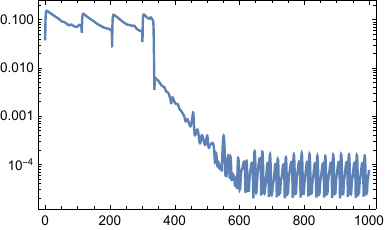}
    };
    \node at (0.25,2.25){\small$\frac{1}{n}\log|\ex|$};
    \node at (0.25,-2.25){\small$t$};
    \end{tikzpicture}
    };
    %%%
    \node at (1.5,-12.5){
    \begin{tikzpicture}
    \node at (0,0){
    \includegraphics[]{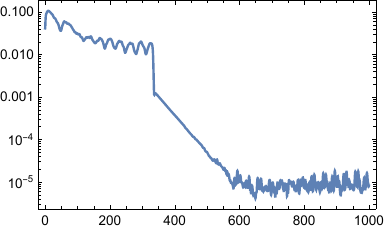}
    };
    \node at (0.25,2.25){\small$\frac{1}{n}\log|\ex|$};
    \node at (0.25,-2.25){\small$t$};
    \end{tikzpicture}
    };
    \node at (9,-12.5){
    \begin{tikzpicture}
    \node at (0,0){
    \includegraphics[]{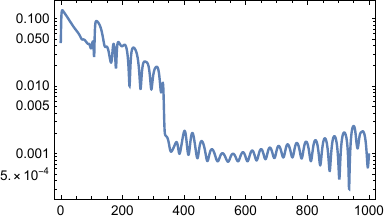}
    };
    \node at (0.25,2.25){\small$\frac{1}{n}\log|\ex|$};
    \node at (0.25,-2.25){\small$t$};
    \end{tikzpicture}
    };
    \node at (4,-5.75) {(a)};
    \node at (11.5,-5.75) {(b)};
    \node at (4,-11) {(c)};
    \node at (11.5,-11) {(d)};
   \end{tikzpicture}
   
    \caption{The first and second rows and panel (a) correspond to a simulation of \eqref{Eq:ClosedLoop_Ideal} with $u_i^*$ is given by \eqref{eq:controlfull} and for $4$ nodes; $x_i$ (plant) is dashed, and $X_i$ (reference) is solid. Panels (a), (b), (c), and (d) show, in logarithmic scale, the mean norm of the error, $\frac{1}{n}|\ex|$: (a) for $4$ nodes, (b) for $100$ nodes, (c) for $100$ nodes with time-varying adjacency matrices, and (d) for $100$ nodes with a mismatch in the reference's parameters used by the controller.
    For $t\in(0,300)$, the controller is off, and hence we see the open-loop response. At $t=300$, the controller is turned on, and from thereon, we observe the closed-loop response. We notice, naturally, that for $t>300$, the plant closely follows the reference. See more details in the main text and compare it with Figure \ref{fig:ideal-hebbian}. }
    \label{fig:u_ideal}
\end{figure*}

A simulation of \eqref{Eq:ClosedLoop_Ideal} with $F_i$ given by \eqref{eq:Fi} is provided in Figure \ref{fig:u_ideal}, recall the general considerations described in Section \ref{sec:generalities_simulations}. Figure \ref{fig:u_ideal} shows in the first two rows and panel (a) the time-series for $x_i(t)$, $X_i(t)$, $u_i(t)$ and the mean norm of the error $\frac{1}{n}|\ex|$ in logarithmic scale for $n=4$ nodes. In this and all following simulations, we prefer to show $\frac{1}{n}|\ex|$ because we keep all the controller constants the same, even for different numbers of nodes. Therefore, for comparison purposes, it is convenient to normalize the error.  In panel (b), we show the mean error for a simulation similar to the previous one but for $n=100$. Panel $(c)$ also shows the mean error for a simulation with $n=100$ nodes but for time-varying adjacency matrices. More precisely, for this simulation, each entry $A_{ij}^\bullet$ (for both the plant and the reference) of the adjacency matrices is of the form $A_{ij}^\bullet=A_{ij}^\bullet(t)=\bar A_{ij}^\bullet(1+\frac{1}{5}\sin(\omega_{ij}t))$ where $\omega_{ij}$ is some random frequency within the interval $(0,1)$ and $\bar A$ is a random rhythmic matrix. Since the controller fully uses these adjacency matrices, its performance is also good. Finally, on panel (d), we show another simulation for $n=100$ nodes, but now with a mismatch between the parameters used by the controller and those of the reference. More precisely, for this simulation, we perturb the parameters $\alpha_r$, $\beta_r$, and $\Ar_{ij}$ used by the controller in \eqref{eq:controlfull}, by some small random number within the interval $(-\frac{1}{20},\frac{1}{20})$. Such small perturbation is different for each parameter and each entry of $\Ar$. We notice that since the mismatch is relatively small, the performance of the controller is still reasonable, although the error is roughly one order of magnitude larger than the error without the mismatch (panel (b)). For all these simulations, we have set $k_i=k=100$. It follows from Proposition \ref{prop:u_ideal} that the larger $k$, the smaller the error.

\subsection{Neuromodulation inspired controller}\label{sec:neuro}

As already mentioned, a drawback of the controller presented in Section \ref{sec:ideal} is its dependence on the full knowledge of the reference's parameters. To offer an alternative, we propose a second controller whose behavior is inspired by the neuromodulation of synaptic weights in neuroscience and neuromorphic systems.

For clarity, let us recall the general error system
\begin{equation}\label{eq:errorr}
    \begin{split}
        \ddt{\ex_i} &= -\ex_i-\ey_i + S\left(\alpha_r \xr_i + \beta_r\sum_{j=1}^n\Ar_{ij}\xr_j\right) \\
        &\quad- S\left(\alpha_p \xp_i + \beta_p\sum_{j=1}^n\Ap_{ij}\xp_j+u_i\right)\\
        \ddt{\ey_i}&=\ve(\ex_i-\ey_i).
    \end{split}
\end{equation}
The controller we now propose is given in an integral form by
\begin{equation}\label{eq:u_hebbian}
\ddt{u_i}=\tau\left(-\lambda u_i+k\ex_i\right),
\end{equation}
with $\tau$, $k$ and $\lambda$, positive constants.

Intuitively, the controller $u_i$ adapts to the rescaled error $\frac{k}{\lambda}\ex_i$ (quickly if $\tau$ is large), and effectively provides a proportional feedback $u_i\approx\frac{k}{\lambda}\ex_i$. If the error $\ex_i$ is large, then the controller compensates the second sigmoid function to balance them out. If the error is small, then both sigmoid functions have roughly the same value and cancel out, which leads to the exponentially stable linear error dynamics \eqref{eq:error0}. Besides the well-known advantages of introducing the controller in integral form, the idea just described has its inspiration in some neurological mechanisms, where the dynamic equation for $u_i$ is in ``leaky form'', similar to the way biological neurons process information. Likewise, the introduced controller can be regarded as an adaptive synaptic weight that modulates the influence of the error. All these interpretations become more evident in Section \ref{sec:implementation}. 
%The term $-\lambda u_i$ in the adaptation rule serves as an intrinsic decay mechanism, which is present in many biological systems, and ensures that $u_i$ remains bounded. 
%Finally, the fact that at steady state $u_i$ is proportional to $\ex_i$ can be interpreted as a Hebbian-like rule where the error $\ex_i$ drives the behavior of $u_i$ (again this will be more evident in Section \ref{sec:implementation}). 

The previous idea is formalized in the following proposition.

\begin{proposition}\label{prop:u_hebbian}
    Consider the error system \eqref{eq:errorr} with the controller given through \eqref{eq:u_hebbian}. If $\tau>0$ and $k>0$ are sufficiently large, then $\lim_{t\to\infty}|\ex(t)|=\cO(\frac{1}{k})$.
\end{proposition}
\begin{proof}
    First, notice that the nonlinear terms in \eqref{eq:errorr} are bounded, that is $|S(\cdot)|\leq m$. Due to the linear part of \eqref{eq:errorr}, its trajectories are globally attracted to a forward invariant set around the origin. Since $\sup_{(a,b)\in\R^2}|S(a)-S(b)|=2m$, a rough estimate of such a region is a ball of radius $2m$ centered at the origin. 
    
    Substituting the steady state value of the controller $u_i=\frac{k}{\lambda}\ex_i=\sigma\ex_i$ into the equation of $\ddt{\ex_i}$ in \eqref{eq:errorr}:
    \begin{equation}
        \begin{split}
            \ddt{\ex_i} &= -\ex_i-\ey_i + S\left(\alpha_r \xr_i + \beta_r\sum_{j=1}^n\Ar_{ij}\xr_j\right) \\
        &- S\left(\alpha_p \xp_i + \beta_p\sum_{j=1}^n\Ap_{ij}\xp_j+\sigma\ex_i\right)\\
        %&=-\ex_i-\ey_i \\
        %&+ S\left(\alpha_r (\ex_i+\xp_i) + \beta_r\sum_{j=1}^n\Ar_{ij} (\ex_j+\xp_j)\right) \\
        %&- S\left(\alpha_p \xp_i + \beta_p\sum_{j=1}^n\Ap_{ij}\xp_j+\sigma\ex_i\right),
        \end{split}
    \end{equation}
    where we recall that $(X,Y)$ as well as $(\xp,\yp)$ are bounded.

    For simplicity let
    \begin{equation}
        \begin{split}
            \zeta_i^r(z) &=\alpha_rz_i+\beta_r\sum_{j=1}^n\Ar_{ij}z_{j},\\
            \zeta_i^p(z)&=\alpha_pz_i+\beta_p\sum_{j=1}^n\Ap_{ij}z_j,
        \end{split}
    \end{equation}
    for $z\in\R^n$. Notice that $\zeta_i^r(z)$ and $\zeta_i^p(z)$, $i=1,\ldots,n$, are bounded along solutions of \eqref{Eq:ClosedLoop_Ideal}, i.e. for $z=x$ or $z=X$.

    Consider the candidate Lyapunov function $V=\frac{1}{2}\sum_{i=1}^n(\ex_i^2+\frac{1}{\ve}\ey_i^2)$. Then
    \begin{equation}
        \begin{split}
            \ddt{V} &= \sum_{i=1}^n\left( -\ex_i^2-\ey_i^2 
            %\right.\\
            %\quad&\left.
            +\ex_i[S(\zeta_i^r(\xr))-S(\zeta^p_i(\xp)+\sigma\ex_i)]\right).
        \end{split}
    \end{equation}
    We notice that $\ex_i S(\zeta_i^r(X))\leq m|\ex_i|$ and that $S(\zeta^p_i(\xp)+\sigma\ex_i)\sim S(\sigma\ex_i)+\cO(\frac{1}{\sigma})\sim m\,\sign(\ex_i)+\cO(\frac{1}{\sigma})$ as $\sigma\to\infty$. This leads to
    \begin{equation}
        \ddt{V}\sim \sum_{i=1}^n\left( -\ex_i^2-\ey_i^2 +\cO\left(\frac{1}{\sigma}\right)\right),
    \end{equation}
    as $\sigma\to\infty$, showing that the region where $\ddt V>0$ is bounded by a ball of radius $\cO\left(\frac{1}{\sigma}\right)$. 
    %Notice that we also see that the larger the network, the larger $\sigma$ is needed to compensate for the number of nodes.
\end{proof}

\begin{figure*}[htbp]
    \centering
    \begin{tikzpicture}
    \node at (0,0){
    \begin{tikzpicture}
    \node at (0,0){
    \includegraphics[]{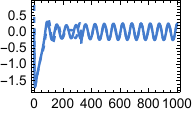}
    };
    \node at (0.5,-1.25) {\small$t$};
    \node[] at (0,1.2) {\small$\xp_1,\xr_1$};
    \end{tikzpicture}
    };
    \node at (3.5,0){
    \begin{tikzpicture}
    \node at (0,0){
    \includegraphics[]{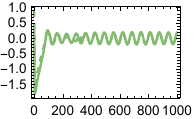}
    };
    \node at (0.5,-1.25) {\small$t$};
    \node[] at (0,1.2) {\small$\xp_2,\xr_2$};
    \end{tikzpicture}
    };
    \node at (7,0){
    \begin{tikzpicture}
    \node at (0,0){
    \includegraphics[]{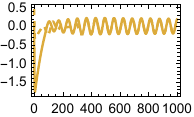}
    };
    \node at (0.5,-1.25) {\small$t$};
    \node[] at (0,1.2) {\small$\xp_3,\xr_3$};
    \end{tikzpicture}
    };
    \node at (10.5,0){
    \begin{tikzpicture}
    \node at (0,0){
    \includegraphics[]{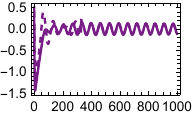}
    };
    \node at (0.5,-1.25) {\small$t$};
    \node[] at (0,1.2) {\small$\xp_4,\xr_4$};
    \end{tikzpicture}
    };
    \node at (0,-3){
    \begin{tikzpicture}
    \node at (0,0){
    \includegraphics[]{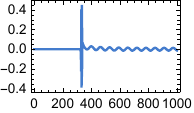}
    };
    \node at (0.5,-1.25) {\small$t$};
    \node[] at (0,1.2) {\small$u_1$};
    \end{tikzpicture}
    };
    \node at (3.5,-3){
    \begin{tikzpicture}
    \node at (0,0){
    \includegraphics[]{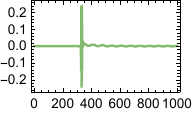}
    };
    \node at (0.5,-1.25) {\small$t$};
    \node[] at (0,1.2) {\small$u_2$};
    \end{tikzpicture}
    };
    \node at (7,-3){
    \begin{tikzpicture}
    \node at (0,0){
    \includegraphics[]{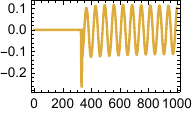}
    };
    \node at (0.5,-1.25) {\small$t$};
    \node[] at (0,1.2) {\small$u_3$};
    \end{tikzpicture}
    };
    \node at (10.5,-3){
    \begin{tikzpicture}
    \node at (0,0){
    \includegraphics[]{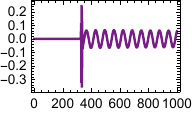}
    };
    \node at (0.5,-1.25) {\small$t$};
    \node[] at (0,1.2) {\small$u_4$};
    \end{tikzpicture}
    };
    %%%
    \node at (1.5,-7.25){
    \begin{tikzpicture}
    \node at (0,0){
    \includegraphics[]{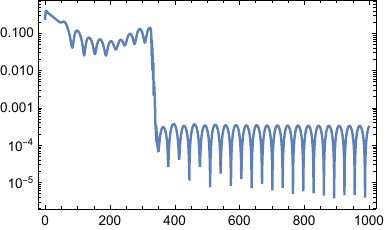}
    };
    \node at (0.25,2.25){\small$\frac{1}{n}\log|\ex|$};
    \node at (0.25,-2.25){\small$t$};
    \end{tikzpicture}
    };
    \node at (9,-7.25){
    \begin{tikzpicture}
    \node at (0,0){
    \includegraphics[]{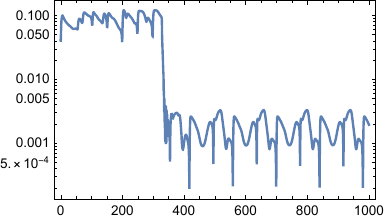}
    };
    \node at (0.25,2.25){\small$\frac{1}{n}\log|\ex|$};
    \node at (0.25,-2.25){\small$t$};
    \end{tikzpicture}
    };
    %%%
    \node at (1.5,-12.5){
    \begin{tikzpicture}
    \node at (0,0){
    \includegraphics[]{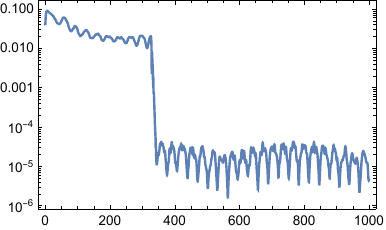}
    };
    \node at (0.25,2.25){\small$\frac{1}{n}\log|\ex|$};
    \node at (0.25,-2.25){\small$t$};
    \end{tikzpicture}
    };
    \node at (4,-5.75) {(a)};
    \node at (11.5,-5.75) {(b)};
    \node at (4,-11) {(c)};
   \end{tikzpicture}
    \caption{Simulation of \eqref{Eq:ClosedLoop_General} with the neuromorphic inspired controller given by \eqref{eq:u_hebbian}. The first and second rows and panel (a) correspond to a simulation with $4$ nodes; $x_i$ (plant) is dashed, and $X_i$ (reference) is solid. Panels (a), (b), and (c) show, in logarithmic scale, the mean norm of the error $\frac{1}{n}|\ex|$: (a) for $4$ nodes, (b) for $100$ nodes, (c) for $100$ nodes with time-varying adjacency matrices.  For $t\in(0,300)$, the controller is off, and hence we see the open-loop response. At $t=300$, the controller is turned on, and from thereon, we observe the closed-loop response. We notice, naturally, that for $t>300$, the plant closely follows the reference. See more details in the main text and compare it with Figure \ref{fig:u_ideal}.}
    \label{fig:ideal-hebbian}
\end{figure*}

Figure \ref{fig:ideal-hebbian} shows a simulation of \eqref{Eq:ClosedLoop_General} with the controller given by \eqref{eq:u_hebbian}, recall the general considerations described in Section \ref{sec:generalities_simulations}. The figure shows in the first two rows and panel (a) the time-series for $x_i(t)$, $X_i(t)$, $u_i(t)$ and the mean norm of the error $\frac{1}{n}|\ex|$ in logarithmic scale for $n=4$ nodes. In panel (b), we show the mean error for a simulation similar to the previous one but for $n=100$. Panel $(c)$ also shows the mean error for a simulation with $n=100$ nodes but for time-varying adjacency matrices in the same way as those for Figure \ref{fig:u_ideal}. Since this controller does not use the parameters of the plant, there is no equivalent simulation to that of the panel (d) in Figure \ref{fig:u_ideal}. For all the simulations in Figure \ref{fig:ideal-hebbian}, $\tau=1$, $\lambda=1$, and $k=100$. It follows from Proposition \ref{prop:u_hebbian} that the larger $k$, the smaller the error.

\section{\label{sec:implementation}Co-evolutionary implementation of the controllers and applications}

In Section \ref{sec:setup}, we have proposed two distinct controllers that render a reference profile stable for a plant with a fixed network structure. A system of the form \eqref{Eq:StaticNetwork} has the potential to be exploited in, for example, neuromorphic applications as it is suitable to be implemented in hardware. Therefore, it is reasonable to consider possible ways to also implement the controllers. While there are many ways to approach this, in this section, we propose to consider an extra node of type \eqref{Eq:StaticNetwork} as the controller. Since the dynamics of the controller's node are relatively simple (either converge to the origin or oscillate) the actual control action is to be performed by adapting the weights of the edges connecting the controller's node to the plant. Biological systems, particularly neuronal ones, are the inspiration for this. The controller is regarded as a pre-synaptic neuron, and the nodes of the plant as post-synaptic, while the adaptation rule is associated with synaptic plasticity.  More precisely, we consider the system
\begin{equation}\label{eq:oneC}
\begin{split}
    \ddt{\xr_i}&=-\xr_i-\yr_i+S\left(\alpha_r \xr_i + \beta_r\sum_{j=1}^n\Ar_{ij}\xr_j\right)\\
    \ddt{\yr_i}&=\ve(\xr_i-\yr_i)\\
    \ddt{\xp_i}&=-\xp_i-\yp_i+S\left(\alpha_p \xp_i + \beta_p\sum_{j=1}^n\Ap_{ij}\xp_j+a_i\xc\right)\\
    \ddt{\yp_i}&=\ve(\xp_i-\yp_i)\\
    \ddt{\xc}&=-\xc-\yc+S\left(\alpha_c \xc\right)\\
    \ddt{\yc}&=\ve(\xc-\yc),\\
    \ddt{a_i}&=\tau_c g_i(a_i,\xc,\yc,\xp,\yp,\ex,\ey).
\end{split}
\end{equation}
where the control is now to be performed by the adaptation rule $g_i$, and $\tau_c>0$ sets the adaptation rate. The parameter $\alpha_c>1$ is chosen so that the controller's node $(x_c,y_c)$ is oscillatory. By ``implementation of the controllers'' we precisely mean that we will choose an adaptation rule $g_i$ to mimic the control actions of those in Sections \ref{sec:ideal} and \ref{sec:neuro}. 

The first controller, given by \eqref{eq:controlfull}, is implemented through the adaptation rule
\begin{equation}\label{eq:ideal_implementation}
    g_i=(a_i^*-a_i),
\end{equation}
where
\begin{equation}\label{eq:ai}
    a_i^*=\begin{cases}
        \frac{F_i}{\xc}, & |\xc|\geq\delta\\
        \frac{F_i}{\delta}, & 0\leq\xc<\delta\\
        -\frac{F_i}{\delta}, & -\delta<\xc<0,
    \end{cases}
\end{equation}
with $0<\delta\ll1$ and $F_i$ given by \eqref{eq:Fi}. Indeed, for $|\xc|>\delta$ we have that $a_i^*\xc=F_i$ and so the effective input to the $i$-th node is exactly as in Proposition \ref{prop:u_ideal}. Since $\xc$ is oscillatory, and crosses the origin, we must account for it and hence propose the piece-wise continuous rule given by \eqref{eq:ai}.

\begin{remark}\leavevmode
\begin{itemize}
    \item The form of \eqref{eq:ai} is due to the multiplicative action of $\xc$. If $\xc=1$, then the controller would be exactly the same as in \eqref{eq:controlfull}. However, from \eqref{eq:mixed}, the only stable equilibrium point of the controller's node dynamics is the origin.
    \item While the adaptation rule is discontinuous, it is reminiscent of sliding mode control \cite{young1999control}. Improvements to this approach, for example, to avoid discontinuities or to rigorously prove the uniqueness of solutions are not further discussed here but shall be considered in future research. 
\end{itemize}
\end{remark}

On the other hand, the controller given in Proposition \ref{prop:u_hebbian} is \emph{heuristically} implemented in \eqref{eq:oneC} by:
\begin{equation}\label{eq:hebbian_implementation}
    g_i=-\lambda a_i + k \ex_i\xc,
\end{equation}
compare with \eqref{eq:u_hebbian}, where we notice that they share the same ``error modulated behavior''. 

Here, the biological inspiration is more evident: the adaptation rule \eqref{eq:hebbian_implementation} has two main components; one is a ``pre-synaptic'' activity-dependent term $\ex_i\xc$, which is regulated by the ``post-synaptic'' error and is reminiscent of a Hebbian-like adaptation (one could say that the learning rate is mediated by how far the plant's node is from the reference's), and an intrinsic decay $-\lambda a_i$ ensuring that the values of $a_i$ remain bounded. 

\begin{remark}
    The adaptation rule \eqref{eq:hebbian_implementation} is \emph{not} obtained by letting $u_i=a_i\xc$ in Proposition \ref{prop:u_hebbian}. In such a case one would obtain
    \begin{equation}\label{eq:ai2}
        \ddt{a_i}=\frac{1}{\xc}\left(  a_i\ddt{\xc}-\tau(-\lambda a_i\xc +  k \ex_i) \right).
    \end{equation}
    While the term $a_i\ddt{\xc}$ in \eqref{eq:ai2} can be interpreted as a ``predictive'' term, we encounter, again, the nuance of dividing by $\xc$. Of course, one could opt to propose a similar discontinuous approach as in the previous case, but this would turn out to be difficult to justify implementation-wise. In addition, \eqref{eq:ai2} lacks an intrinsic term guaranteeing boundedness, which is important in the present context.
    
    A more sensible perspective is to recall that \eqref{eq:u_hebbian} effectively induces the feedback $u_i=k\ex_i$ into the error system. If now $u_i$ is considered as $u_i=a_i\xc$, then the ideal value of $a_i$ would be $a_i=\frac{k\ex_i}{\xc}$, again introducing the division by $\xc$. In our heuristic implementation, we consider $x_c$ as a scaling, or gain, being its sign its most important aspect leading to \eqref{eq:hebbian_implementation}. Naturally, since $\xc$ crosses zero, each time $\xc=0$, the ``learned'' value of $a_i$ will be forgotten. This is indeed visible in the simulations presented below.
\end{remark}

In the following subsections, we present simulations of \eqref{eq:oneC}, and we compare through different application scenarios, the adaptation rules given by \eqref{eq:ideal_implementation}-\eqref{eq:ai} and \eqref{eq:hebbian_implementation}. In addition to the general considerations for our numerical simulations described in Section \ref{sec:generalities_simulations}, we add the following: in the time interval $t\in(500,550)$ we add a constant random disturbance, taking values in the interval $(-1,1)$, to each node; different for each node and independent of the node. We do this to numerically test how the controlled plant recovers from a perturbed state.

To avoid repetitions, we mention here once that in our numerical simulations of \eqref{eq:oneC} with the adaptive implementation given by \eqref{eq:ideal_implementation}-\eqref{eq:ai} (specifically Figures \ref{fig:ideal_synch}, \ref{fig:arbitrary_ideal} and \ref{fig:tv_ideal}) we have used $\delta=\frac{1}{100}$ and $\tau_c=100$. Regarding the adaptive implementation given by \eqref{eq:hebbian_implementation} (specifically Figures \ref{fig:hebbian_synch}, \ref{fig:arbitrary_hebbian} and \ref{fig:tv_hebbian}) we have used $\tau_c=100$, $k=100$ and $\lambda=1$. We have kept these values the same across the different simulations for comparison purposes; of course, the performance of the adaptation rules can be improved if one tunes such parameters adequately.

\subsection{Synchronization}\label{sec:synch}

In this section, we consider the problem of synchronization of the plant with respect to a provided reference. For this, the reference network has one node $(\xr_1,\yr_1)$, and the errors are thus defined by
\begin{equation}
    \ex_i=\xr_1-\xp_i,\qquad \ey_i=\yr_1-\yp_i.
\end{equation}

\begin{figure*}[htbp]
    \centering
    \begin{tikzpicture}
    \node at (0,0){
    \begin{tikzpicture}
    \node at (0,0){
    \includegraphics[]{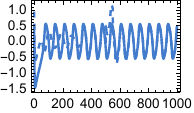}
    };
    \node at (0.5,-1.25) {\small$t$};
    \node[] at (0,1.2) {\small$\xp_1,\xr_1$};
    \end{tikzpicture}
    };
    \node at (3.5,0){
    \begin{tikzpicture}
    \node at (0,0){
    \includegraphics[]{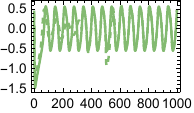}
    };
    \node at (0.5,-1.25) {\small$t$};
    \node[] at (0,1.2) {\small$\xp_2,\xr_1$};
    \end{tikzpicture}
    };
    \node at (7,0){
    \begin{tikzpicture}
    \node at (0,0){
    \includegraphics[]{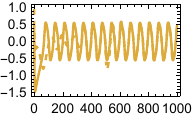}
    };
    \node at (0.5,-1.25) {\small$t$};
    \node[] at (0,1.2) {\small$\xp_3,\xr_1$};
    \end{tikzpicture}
    };
    \node at (10.5,0){
    \begin{tikzpicture}
    \node at (0,0){
    \includegraphics[]{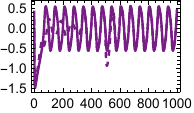}
    };
    \node at (0.5,-1.25) {\small$t$};
    \node[] at (0,1.2) {\small$\xp_4,\xr_1$};
    \end{tikzpicture}
    };
    \node at (0,-3){
    \begin{tikzpicture}
    \node at (0,0){
    \includegraphics[]{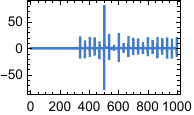}
    };
    \node at (0.5,-1.25) {\small$t$};
    \node[] at (0,1.2) {\small$a_1$};
    \end{tikzpicture}
    };
    \node at (3.5,-3){
    \begin{tikzpicture}
    \node at (0,0){
    \includegraphics[]{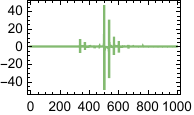}
    };
    \node at (0.5,-1.25) {\small$t$};
    \node[] at (0,1.2) {\small$a_2$};
    \end{tikzpicture}
    };
    \node at (7,-3){
    \begin{tikzpicture}
    \node at (0,0){
    \includegraphics[]{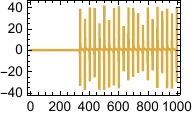}
    };
    \node at (0.5,-1.25) {\small$t$};
    \node[] at (0,1.2) {\small$a_3$};
    \end{tikzpicture}
    };
    \node at (10.5,-3){
    \begin{tikzpicture}
    \node at (0,0){
    \includegraphics[]{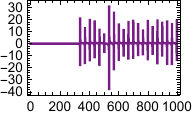}
    };
    \node at (0.5,-1.25) {\small$t$};
    \node[] at (0,1.2) {\small$a_4$};
    \end{tikzpicture}
    };
    %%%
    \node at (1.5,-7.25){
    \begin{tikzpicture}
    \node at (0,0){
    \includegraphics[]{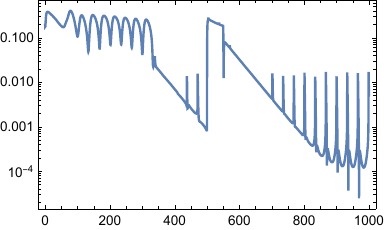}
    };
    \node at (0.25,2.25){\small$\frac{1}{n}\log|\ex|$};
    \node at (0.25,-2.25){\small$t$};
    \end{tikzpicture}
    };
    \node at (9,-7.25){
    \begin{tikzpicture}
    \node at (0,0){
    \includegraphics[]{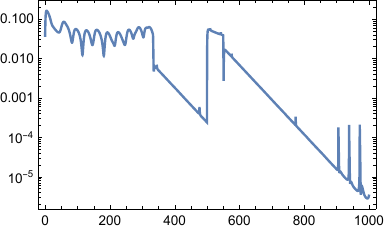}
    };
    \node at (0.25,2.25){\small$\frac{1}{n}\log|\ex|$};
    \node at (0.25,-2.25){\small$t$};
    \end{tikzpicture}
    };
    \node at (4,-5.75) {(a)};
    \node at (11.5,-5.75) {(b)};
   \end{tikzpicture}
   
    \caption{Synchronization of the plant with respect to a reference node using the adaptation rule \eqref{eq:ideal_implementation}-\eqref{eq:ai}. The first and second rows show the time series of $X_1$ (solid) and $x_i$ (dashed) and the corresponding adaptive weights $a_i$ for $n=4$ nodes while panel (a) shows the corresponding error in logarithmic scale. Panel (b) shows a similar simulation but for $n=100$ nodes. In all simulations, we first show the system in open-loop for $t\in[0,300)$. At $t=300$ the controller is turned on, and we see how the error decreases. For $t\in[500,550]$ a random constant disturbance is added to each node of the plant, hence the observed increase in the error.  After $t=550$ we see how the error again decreases, showcasing the recovery from the added disturbance.}
    \label{fig:ideal_synch}
\end{figure*}

\begin{figure*}[htbp]
    \centering
    \begin{tikzpicture}
    \node at (0,0){
    \begin{tikzpicture}
    \node at (0,0){
    \includegraphics[]{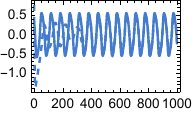}
    };
    \node at (0.5,-1.25) {\small$t$};
    \node[] at (0,1.2) {\small$\xp_1,\xr_1$};
    \end{tikzpicture}
    };
    \node at (3.5,0){
    \begin{tikzpicture}
    \node at (0,0){
    \includegraphics[]{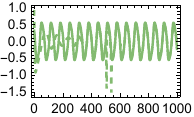}
    };
    \node at (0.5,-1.25) {\small$t$};
    \node[] at (0,1.2) {\small$\xp_2,\xr_2$};
    \end{tikzpicture}
    };
    \node at (7,0){
    \begin{tikzpicture}
    \node at (0,0){
    \includegraphics[]{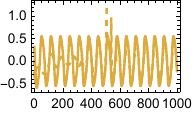}
    };
    \node at (0.5,-1.25) {\small$t$};
    \node[] at (0,1.2) {\small$\xp_3,\xr_3$};
    \end{tikzpicture}
    };
    \node at (10.5,0){
    \begin{tikzpicture}
    \node at (0,0){
    \includegraphics[]{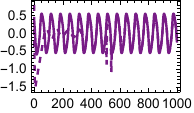}
    };
    \node at (0.5,-1.25) {\small$t$};
    \node[] at (0,1.2) {\small$\xp_4,\xr_4$};
    \end{tikzpicture}
    };
    \node at (0,-3){
    \begin{tikzpicture}
    \node at (0,0){
    \includegraphics[]{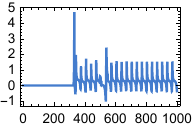}
    };
    \node at (0.5,-1.25) {\small$t$};
    \node[] at (0,1.2) {\small$a_1$};
    \end{tikzpicture}
    };
    \node at (3.5,-3){
    \begin{tikzpicture}
    \node at (0,0){
    \includegraphics[]{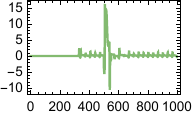}
    };
    \node at (0.5,-1.25) {\small$t$};
    \node[] at (0,1.2) {\small$a_2$};
    \end{tikzpicture}
    };
    \node at (7,-3){
    \begin{tikzpicture}
    \node at (0,0){
    \includegraphics[]{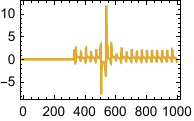}
    };
    \node at (0.5,-1.25) {\small$t$};
    \node[] at (0,1.2) {\small$a_3$};
    \end{tikzpicture}
    };
    \node at (10.5,-3){
    \begin{tikzpicture}
    \node at (0,0){
    \includegraphics[]{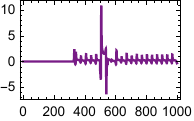}
    };
    \node at (0.5,-1.25) {\small$t$};
    \node[] at (0,1.2) {\small$a_4$};
    \end{tikzpicture}
    };
    %%%
    \node at (1.5,-7.25){
    \begin{tikzpicture}
    \node at (0,0){
    \includegraphics[]{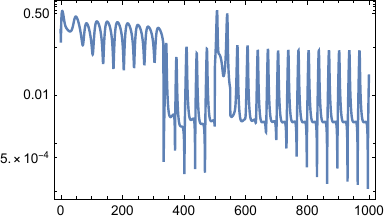}
    };
    \node at (0.25,2.25){\small$\frac{1}{n}\log|\ex|$};
    \node at (0.25,-2.25){\small$t$};
    \end{tikzpicture}
    };
    \node at (9,-7.25){
    \begin{tikzpicture}
    \node at (0,0){
    \includegraphics[]{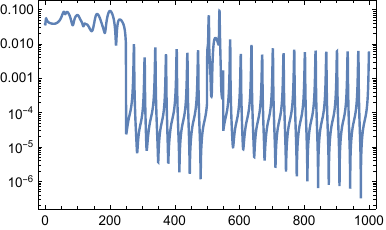}
    };
    \node at (0.25,2.25){\small$\frac{1}{n}\log|\ex|$};
    \node at (0.25,-2.25){\small$t$};
    \end{tikzpicture}
    };
   \end{tikzpicture}
   
    \caption{Synchronization of the plant with respect to a reference node using the Hebbian-like adaptation rule \eqref{eq:hebbian_implementation}. The first and second rows show the time series of $X_1$ (solid) and $x_i$ (dashed) and the adaptive weights $a_i$ for $n=4$ nodes while panel (a) shows the corresponding error in logarithmic scale. Panel (b) shows a similar simulation but for $n=100$ nodes. In all simulations, we first show the system in open-loop for $t\in[0,300)$. At $t=300$ the controller is turned on, and we see how the error decreases. For $t\in[500,550]$, a random constant disturbance is added to each node of the plant, hence the observed increase in the error.  After $t=550$, we see how the error again decreases, showcasing the recovery from the added disturbance. The most important distinction with Figure \ref{fig:ideal_synch} is that we observe periodic increases of the error, which coincide with the crossings of $\xc$ at the origin and due to the multiplicative effect of $\xc$ in the interconnection, namely the term $a_i\xc$ in each equation of the nodes.}
    \label{fig:hebbian_synch}
\end{figure*}

Figure \ref{fig:ideal_synch} shows a simulation of \eqref{eq:oneC} with the adaptation rule given by \eqref{eq:ideal_implementation}-\eqref{eq:ai}. The observed spikes in the weight $a_i$ are due to the discontinuous form of the adaptation and are modulated by $\delta$ in \eqref{eq:ai}. In contrast, Figure \ref{fig:hebbian_synch} shows a simulation of \eqref{eq:oneC} with the adaptation rule given by \eqref{eq:hebbian_implementation}. For this case, we do not see anymore the aforementioned spiking behavior since $a_i$ is smooth. However, we observe that the error tends to increase periodically, and this coincides with $\xc$ crossing zero, as for $\xc=0$ the system is in open-loop. This issue is barely noticeable in the time-series, though. We moreover notice that on both adaptation rules, the closed-loop systems recover well enough after the added disturbance in $t\in[500,550]$.

\begin{remark}
    While our main exposition has considered that the reference and the plant have the same number of nodes, it is clear that, as in this section, such a consideration is not necessary. If the plant has $n\geq1$ nodes and the reference has $m\geq1$ nodes, one simply needs to define $n$ errors $\ex_i=\xr_j-\xp_i$ for $i=1,\ldots,n$ and $j\in\left\{1,\ldots,m \right\}$.
\end{remark}

\subsection{Arbitrary network topologies}\label{sec:arbitrary}

\begin{figure*}[htbp]
    \centering
    \begin{tikzpicture}
    \node at (2,3){
    \includegraphics[]{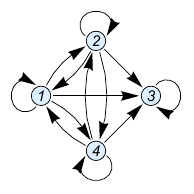}
    };
    \node at (8,3){
    \includegraphics[]{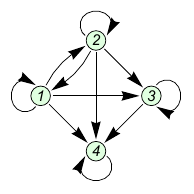}
    };
    
    \node at (0,0){
    \begin{tikzpicture}
    \node at (0,0){
    \includegraphics[]{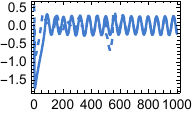}
    };
    \node at (0.5,-1.25) {\small$t$};
    \node[] at (0,1.2) {\small$\xp_1,\xr_1$};
    \end{tikzpicture}
    };
    \node at (3.5,0){
    \begin{tikzpicture}
    \node at (0,0){
    \includegraphics[]{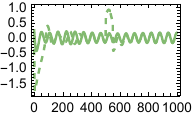}
    };
    \node at (0.5,-1.25) {\small$t$};
    \node[] at (0,1.2) {\small$\xp_2,\xr_2$};
    \end{tikzpicture}
    };
    \node at (7,0){
    \begin{tikzpicture}
    \node at (0,0){
    \includegraphics[]{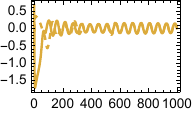}
    };
    \node at (0.5,-1.25) {\small$t$};
    \node[] at (0,1.2) {\small$\xp_3,\xr_3$};
    \end{tikzpicture}
    };
    \node at (10.5,0){
    \begin{tikzpicture}
    \node at (0,0){
    \includegraphics[]{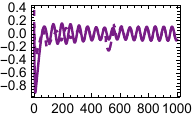}
    };
    \node at (0.5,-1.25) {\small$t$};
    \node[] at (0,1.2) {\small$\xp_4,\xr_4$};
    \end{tikzpicture}
    };
    \node at (0,-3){
    \begin{tikzpicture}
    \node at (0,0){
    \includegraphics[]{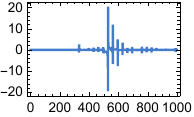}
    };
    \node at (0.5,-1.25) {\small$t$};
    \node[] at (0,1.2) {\small$a_1$};
    \end{tikzpicture}
    };
    \node at (3.5,-3){
    \begin{tikzpicture}
    \node at (0,0){
    \includegraphics[]{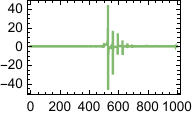}
    };
    \node at (0.5,-1.25) {\small$t$};
    \node[] at (0,1.2) {\small$a_2$};
    \end{tikzpicture}
    };
    \node at (7,-3){
    \begin{tikzpicture}
    \node at (0,0){
    \includegraphics[]{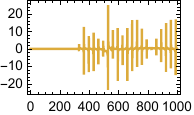}
    };
    \node at (0.5,-1.25) {\small$t$};
    \node[] at (0,1.2) {\small$a_3$};
    \end{tikzpicture}
    };
    \node at (10.5,-3){
    \begin{tikzpicture}
    \node at (0,0){
    \includegraphics[]{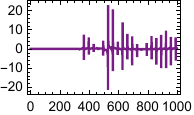}
    };
    \node at (0.5,-1.25) {\small$t$};
    \node[] at (0,1.2) {\small$a_4$};
    \end{tikzpicture}
    };
    %%%
    \node at (1.5,-7.25){
    \begin{tikzpicture}
    \node at (0,0){
    \includegraphics[]{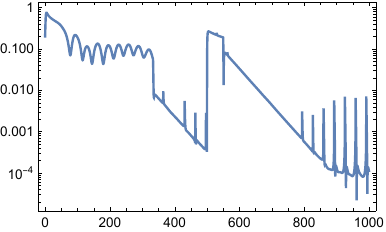}
    };
    \node at (0.25,2.25){\small$\frac{1}{n}\log|\ex|$};
    \node at (0.25,-2.25){\small$t$};
    \end{tikzpicture}
    };
    \node at (9,-7.25){
    \begin{tikzpicture}
    \node at (0,0){
    \includegraphics[]{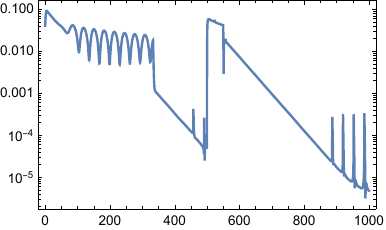}
    };
    \node at (0.25,2.25){\small$\frac{1}{n}\log|\ex|$};
    \node at (0.25,-2.25){\small$t$};
    \end{tikzpicture}
    };
   \end{tikzpicture}
   
    \caption{Simulation of \eqref{eq:oneC} with \eqref{eq:ideal_implementation}-\eqref{eq:ai} where the plant and the reference have different network topologies. At the top of the figure, we show on the left the reference's graph and on the right the plant's graph for $n=4$, notice that they are indeed different. The second and third rows show the time series of $X_1$ (solid) and $x_i$ (dashed) and the adaptive weights $a_i$ for $n=4$ nodes while panel (a) shows the corresponding error in logarithmic scale. Panel (b) shows a similar simulation but for $n=100$ nodes. In all simulations, we first show the system in open-loop for $t\in[0,300)$. At $t=300$ the controller is turned on, and we see how the error decreases. For $t\in[500,550]$, a random constant disturbance is added to each node of the plant, hence the observed increase in the error.  After $t=550$ we see how the error again decreases, showcasing the recovery from the added disturbance.}
    \label{fig:arbitrary_ideal}
\end{figure*}

\begin{figure*}[htbp]
    \centering
    \begin{tikzpicture}
    \node at (2,3){
    \includegraphics[]{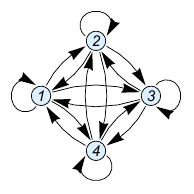}
    };
    \node at (8,3){
    \includegraphics[]{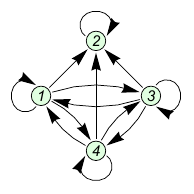}
    };
    \node at (0,0){
    \begin{tikzpicture}
    \node at (0,0){
    \includegraphics[]{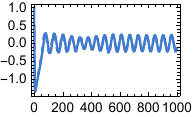}
    };
    \node at (0.5,-1.25) {\small$t$};
    \node[] at (0,1.2) {\small$\xp_1,\xr_1$};
    \end{tikzpicture}
    };
    \node at (3.5,0){
    \begin{tikzpicture}
    \node at (0,0){
    \includegraphics[]{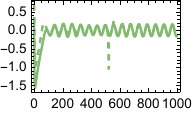}
    };
    \node at (0.5,-1.25) {\small$t$};
    \node[] at (0,1.2) {\small$\xp_2,\xr_2$};
    \end{tikzpicture}
    };
    \node at (7,0){
    \begin{tikzpicture}
    \node at (0,0){
    \includegraphics[]{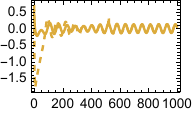}
    };
    \node at (0.5,-1.25) {\small$t$};
    \node[] at (0,1.2) {\small$\xp_3,\xr_3$};
    \end{tikzpicture}
    };
    \node at (10.5,0){
    \begin{tikzpicture}
    \node at (0,0){
    \includegraphics[]{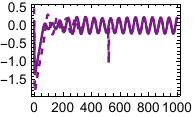}
    };
    \node at (0.5,-1.25) {\small$t$};
    \node[] at (0,1.2) {\small$\xp_4,\xr_4$};
    \end{tikzpicture}
    };
    \node at (0,-3){
    \begin{tikzpicture}
    \node at (0,0){
    \includegraphics[]{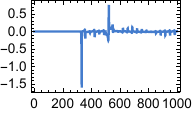}
    };
    \node at (0.5,-1.25) {\small$t$};
    \node[] at (0,1.2) {\small$a_1$};
    \end{tikzpicture}
    };
    \node at (3.5,-3){
    \begin{tikzpicture}
    \node at (0,0){
    \includegraphics[]{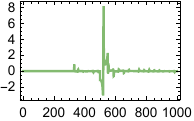}
    };
    \node at (0.5,-1.25) {\small$t$};
    \node[] at (0,1.2) {\small$a_2$};
    \end{tikzpicture}
    };
    \node at (7,-3){
    \begin{tikzpicture}
    \node at (0,0){
    \includegraphics[]{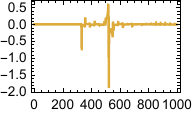}
    };
    \node at (0.5,-1.25) {\small$t$};
    \node[] at (0,1.2) {\small$a_3$};
    \end{tikzpicture}
    };
    \node at (10.5,-3){
    \begin{tikzpicture}
    \node at (0,0){
    \includegraphics[]{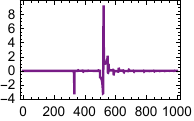}
    };
    \node at (0.5,-1.25) {\small$t$};
    \node[] at (0,1.2) {\small$a_4$};
    \end{tikzpicture}
    };
    %%%
    \node at (1.5,-7.25){
    \begin{tikzpicture}
    \node at (0,0){
    \includegraphics[]{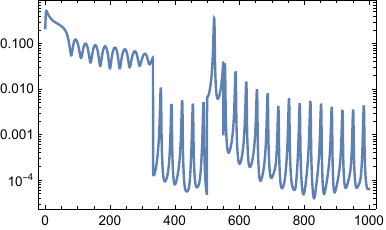}
    };
    \node at (0.25,2.25){\small$\frac{1}{n}\log|\ex|$};
    \node at (0.25,-2.25){\small$t$};
    \end{tikzpicture}
    };
    \node at (9,-7.25){
    \begin{tikzpicture}
    \node at (0,0){
    \includegraphics[]{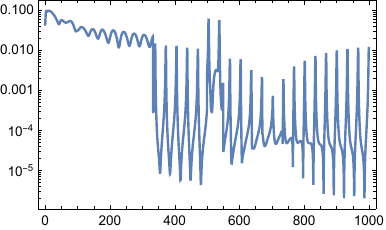}
    };
    \node at (0.25,2.25){\small$\frac{1}{n}\log|\ex|$};
    \node at (0.25,-2.25){\small$t$};
    \end{tikzpicture}
    };
   \end{tikzpicture}
   
    \caption{Simulation of \eqref{eq:oneC} with \eqref{eq:hebbian_implementation} where the plant and the reference have different network topologies. At the top of the figure, we show on the left the reference's graph and on the right the plant's graph for $n=4$, notice that they are indeed different. The second and third rows show the time series of $X_1$ (solid) and $x_i$ (dashed) and the adaptive weights $a_i$ for $n=4$ nodes while panel (a) shows the corresponding error in logarithmic scale. Panel (b) shows a similar simulation but for $n=100$ nodes. In all simulations, we first show the system in open-loop for $t\in[0,300)$. At $t=300$ the controller is turned on, and we see how the error decreases. For $t\in[500,550]$, a random constant disturbance is added to each node of the plant, hence the observed increase in the error.  After $t=550$ we see how the error again decreases, showcasing the recovery from the added disturbance.}
    \label{fig:arbitrary_hebbian}
\end{figure*}

To highlight the fact that our approach does not depend on the particular topologies of the reference's and plant's adjacency matrices, in this section, we present simulations with explicitly different adjacency matrices. Figure \ref{fig:arbitrary_ideal} shows a simulation of \eqref{eq:ideal_implementation}-\eqref{eq:ai} where, for $n=4$, we have included the graphs of the reference and of the plant, which are different. As in the simulations of the previous section, the spikes in the time series of $a_i$ are due to the discontinuous implementation \eqref{eq:ai} and are modulated by $\delta$ in \eqref{eq:ai}. For comparison, Figure \ref{fig:arbitrary_hebbian} shows a simulation of \eqref{eq:hebbian_implementation}. Both simulations follow the same overall format we have used so far: for $t\in[0,300]$, the controller is off, and hence we see the open-loop response; at $t=300$ the controller is turned on, and we see how the plant follows the reference while for $t\in[500,550]$ a constant random disturbance is added to each node. We notice from the time-series that the performance of both controllers is reasonably good.

\subsection{Time-varying adjacency matrices}

\begin{figure*}[htbp]
    \centering
    \begin{tikzpicture}
    \node at (0,0){
    \begin{tikzpicture}
    \node at (0,0){
    \includegraphics[]{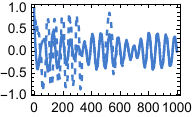}
    };
    \node at (0.5,-1.25) {\small$t$};
    \node[] at (0,1.2) {\small$\xp_1,\xr_1$};
    \end{tikzpicture}
    };
    \node at (3.5,0){
    \begin{tikzpicture}
    \node at (0,0){
    \includegraphics[]{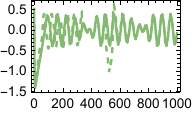}
    };
    \node at (0.5,-1.25) {\small$t$};
    \node[] at (0,1.2) {\small$\xp_2,\xr_2$};
    \end{tikzpicture}
    };
    \node at (7,0){
    \begin{tikzpicture}
    \node at (0,0){
    \includegraphics[]{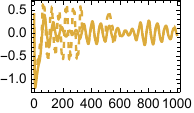}
    };
    \node at (0.5,-1.25) {\small$t$};
    \node[] at (0,1.2) {\small$\xp_3,\xr_3$};
    \end{tikzpicture}
    };
    \node at (10.5,0){
    \begin{tikzpicture}
    \node at (0,0){
    \includegraphics[]{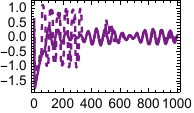}
    };
    \node at (0.5,-1.25) {\small$t$};
    \node[] at (0,1.2) {\small$\xp_4,\xr_4$};
    \end{tikzpicture}
    };
    \node at (0,-3){
    \begin{tikzpicture}
    \node at (0,0){
    \includegraphics[]{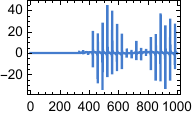}
    };
    \node at (0.5,-1.25) {\small$t$};
    \node[] at (0,1.2) {\small$a_1$};
    \end{tikzpicture}
    };
    \node at (3.5,-3){
    \begin{tikzpicture}
    \node at (0,0){
    \includegraphics[]{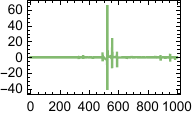}
    };
    \node at (0.5,-1.25) {\small$t$};
    \node[] at (0,1.2) {\small$a_2$};
    \end{tikzpicture}
    };
    \node at (7,-3){
    \begin{tikzpicture}
    \node at (0,0){
    \includegraphics[]{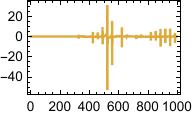}
    };
    \node at (0.5,-1.25) {\small$t$};
    \node[] at (0,1.2) {\small$a_3$};
    \end{tikzpicture}
    };
    \node at (10.5,-3){
    \begin{tikzpicture}
    \node at (0,0){
    \includegraphics[]{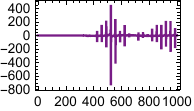}
    };
    \node at (0.5,-1.25) {\small$t$};
    \node[] at (0,1.2) {\small$a_4$};
    \end{tikzpicture}
    };
    %%%
    \node at (1.5,-7.25){
    \begin{tikzpicture}
    \node at (0,0){
    \includegraphics[]{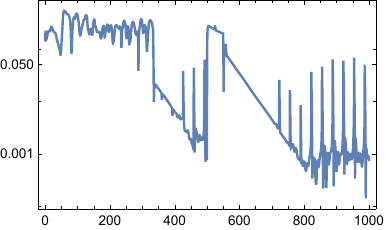}
    };
    \node at (0.25,2.25){\small$\frac{1}{n}\log|\ex|$};
    \node at (0.25,-2.25){\small$t$};
    \end{tikzpicture}
    };
    \node at (9,-7.25){
    \begin{tikzpicture}
    \node at (0,0){
    \includegraphics[]{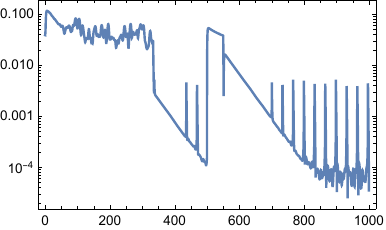}
    };
    \node at (0.25,2.25){\small$\frac{1}{n}\log|\ex|$};
    \node at (0.25,-2.25){\small$t$};
    \end{tikzpicture}
    };
   \end{tikzpicture}
   
    \caption{Simulation of \eqref{eq:oneC} with \eqref{eq:ideal_implementation}-\eqref{eq:ai} where the plant and the reference have different network topologies with time-varying weights. The first and second rows show the time series of $X_1$ (solid) and $x_i$ (dashed) and the adaptive weights $a_i$ for $n=4$ nodes while panel (a) shows the corresponding error in logarithmic scale. Panel (b) shows a similar simulation but for $n=100$ nodes. In all simulations, we first show the system in open-loop for $t\in[0,300)$. At $t=300$, the controller is turned on, and we see how the error decreases. For $t\in[500,550]$, a random constant disturbance is added to each node of the plant, hence the observed increase in the error.  After $t=550$, we see how the error again decreases, showcasing the recovery from the added disturbance.}
    \label{fig:tv_ideal}
\end{figure*}

\begin{figure*}[htbp]
    \centering
    \begin{tikzpicture}
    \node at (0,0){
    \begin{tikzpicture}
    \node at (0,0){
    \includegraphics[]{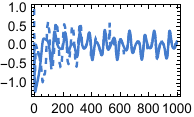}
    };
    \node at (0.5,-1.25) {\small$t$};
    \node[] at (0,1.2) {\small$\xp_1,\xr_1$};
    \end{tikzpicture}
    };
    \node at (3.5,0){
    \begin{tikzpicture}
    \node at (0,0){
    \includegraphics[]{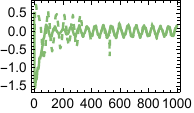}
    };
    \node at (0.5,-1.25) {\small$t$};
    \node[] at (0,1.2) {\small$\xp_2,\xr_2$};
    \end{tikzpicture}
    };
    \node at (7,0){
    \begin{tikzpicture}
    \node at (0,0){
    \includegraphics[]{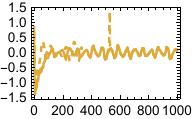}
    };
    \node at (0.5,-1.25) {\small$t$};
    \node[] at (0,1.2) {\small$\xp_3,\xr_3$};
    \end{tikzpicture}
    };
    \node at (10.5,0){
    \begin{tikzpicture}
    \node at (0,0){
    \includegraphics[]{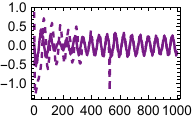}
    };
    \node at (0.5,-1.25) {\small$t$};
    \node[] at (0,1.2) {\small$\xp_4,\xr_4$};
    \end{tikzpicture}
    };
    \node at (0,-3){
    \begin{tikzpicture}
    \node at (0,0){
    \includegraphics[]{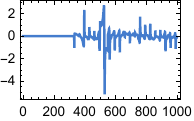}
    };
    \node at (0.5,-1.25) {\small$t$};
    \node[] at (0,1.2) {\small$a_1$};
    \end{tikzpicture}
    };
    \node at (3.5,-3){
    \begin{tikzpicture}
    \node at (0,0){
    \includegraphics[]{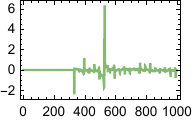}
    };
    \node at (0.5,-1.25) {\small$t$};
    \node[] at (0,1.2) {\small$a_2$};
    \end{tikzpicture}
    };
    \node at (7,-3){
    \begin{tikzpicture}
    \node at (0,0){
    \includegraphics[]{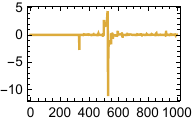}
    };
    \node at (0.5,-1.25) {\small$t$};
    \node[] at (0,1.2) {\small$a_3$};
    \end{tikzpicture}
    };
    \node at (10.5,-3){
    \begin{tikzpicture}
    \node at (0,0){
    \includegraphics[]{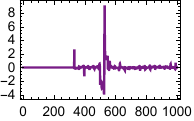}
    };
    \node at (0.5,-1.25) {\small$t$};
    \node[] at (0,1.2) {\small$a_4$};
    \end{tikzpicture}
    };
    %%%
    \node at (1.5,-7.25){
    \begin{tikzpicture}
    \node at (0,0){
    \includegraphics[]{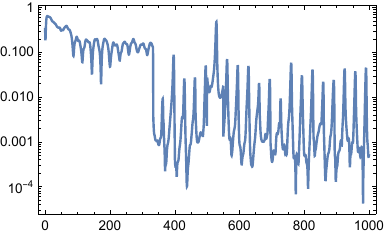}
    };
    \node at (0.25,2.25){\small$\frac{1}{n}\log|\ex|$};
    \node at (0.25,-2.25){\small$t$};
    \end{tikzpicture}
    };
    \node at (9,-7.25){
    \begin{tikzpicture}
    \node at (0,0){
    \includegraphics[]{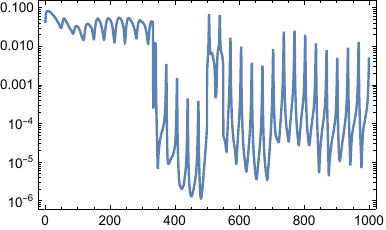}
    };
    \node at (0.25,2.25){\small$\frac{1}{n}\log|\ex|$};
    \node at (0.25,-2.25){\small$t$};
    \end{tikzpicture}
    };
   \end{tikzpicture}
   
    \caption{Simulation of \eqref{eq:oneC} with \eqref{eq:hebbian_implementation} where the plant and the reference have different network topologies with time-varying weights. The first and second rows show the time series of $X_1$ (solid) and $x_i$ (dashed) and the adaptive weights $a_i$ for $n=4$ nodes while panel (a) shows the corresponding error in logarithmic scale. Panel (b) shows a similar simulation but for $n=100$ nodes. In all simulations, we first show the system in open-loop for $t\in[0,300)$. At $t=300$, the controller is turned on, and we see how the error decreases. For $t\in[500,550]$, a random constant disturbance is added to each node of the plant, hence the observed increase in the error.  After $t=550$, we see how the error again decreases, showcasing the recovery from the added disturbance.}
    \label{fig:tv_hebbian}
\end{figure*}

As a final application and comparison setup, we consider here that the adjacency matrices of both the reference and the plant are time-varying. We do this similar to what was presented in Section \ref{sec:setup}. More precisely, for the simulations of this section, each entry $A_{ij}^\bullet$ (for both the plant and the reference) of the adjacency matrices is of the form $A_{ij}^\bullet=A_{ij}^\bullet(t)=\bar A_{ij}^\bullet(1+\frac{1}{5}\sin(\omega_{ij}t))$ where $\omega_{ij}$ is some random frequency within the interval $(0,1)$ and $\bar A^{\bullet}$ is a random rhythmic matrix, meaning that the underlying topologies of each network are also different. While we do not perform formal analysis under this setup, both adaptation approaches seem suitable to handle this scenario. On the one hand, we have that the adaptation rule \eqref{eq:ideal_implementation}-\eqref{eq:ai} uses the full knowledge of the adjacency matrices. On the other hand, the adaptive rule \eqref{eq:hebbian_implementation} accounts for the time-varying adjacency matrices from the fact that $a_i$ is regulated by the error. Figure \ref{fig:tv_ideal} corresponds to \eqref{eq:oneC} with \eqref{eq:ideal_implementation}-\eqref{eq:ai}, while Figure \ref{fig:tv_hebbian} uses \eqref{eq:hebbian_implementation}. The description of the results is similar to the previous sections. In both cases we can observe that, despite the adjacency matrices being time-varying, both adaptation rules perform well.

\section{\label{sec:conclusions}Conclusion and discussion}

This paper has explored co-evolutionary control approaches to render a desired oscillatory pattern stable for a particular class of dynamic networks, keeping their internal properties fixed. We have developed two kinds of controllers: one that uses the full information from the reference and another that only requires local error information. From our analysis, both controllers seem to perform reasonably well. Their main difference is their potential implementation. While the first controller (see Proposition \ref{prop:u_ideal}) performs observably well, it is hardly implementable not only because it requires full information about the plant but also because of its particular form (see \eqref{eq:Fi}). In contrast, the second controller (see Proposition \ref{prop:u_hebbian}) requires only local error information. We have also implemented such controllers as co-evolutionary rules in an extended dynamic network and tested their performance in three relevant scenarios. In all cases, again, the controllers perform reasonably well.

An immediate open problem to consider in the future is to extend our approach to more general coupled systems. As a concrete example, one may want to explore ideas similar to those developed here, but for coupled spiking neurons, whose models also have mixed feedback loops. A challenge for these models is that the nonlinearity is not bounded, in contrast to \eqref{eq:mixed}. Another possibility is to consider that the reference and the plant are not of the same type. Hence, even if a system cannot inherently produce and sustain specific oscillations, a controller may impose them.

\begin{acknowledgments}

The work of LGVP was founded by the Data Science and Systems Complexity (DSSC) Centre of the University of Groningen, as part of a PhD scholarship.

HJK would like to acknowledge the financial support of the CogniGron research center and the Ubbo Emmius Funds (Univ. of Groningen).

The authors thank the anonymous reviewers for their thorough comments that helped to improve the manuscript.
\end{acknowledgments}

   \section*{Author Declarations}
        \subsection*{Conflict of Interest}
            The authors have no conflicts to disclose.
        \subsection*{Author Contributions}
            \textbf{Luis Guillermo Venegas-Pineda:} Conceptualization; Methodology; Software; Formal analysis; Data curation; Visualization; Writing - original draft; Writing/Review \& Editing. \textbf{Hildeberto Jardón-Kojakhmetov:} Conceptualization; Methodology; Software; Formal analysis; Writing - original draft; Writing/Review \& Editing; Supervision; Project administration. \textbf{Ming Cao:} Writing/Review \& Editing;  Supervision; Project administration.

\section*{Data Availability Statement}
All codes used in the paper are available in \cite{Jardon-Kojakhmetov_Code_for_Co-evolutionary_2025}.

%\nocite{*}
\bibliography{bibliography}% Produces the bibliography via BibTeX.

\end{document}